\documentclass{article}
\usepackage{graphicx,amsmath}

\newcommand{\bea}{\begin{eqnarray}}
\newcommand{\eea}{\end{eqnarray}}
\newcommand{\be}{\begin{equation}}
\newcommand{\ee}{\end{equation}}

\def\be{\begin{eqnarray}}
\def\ee{\end{eqnarray}}
\def\bd{\begin{displaymath}}
\def\ed{\end{displaymath}}

\def\ga{\gamma}

\def\etal{{\em et al.}}
\def\ADNDT{{\em At. Data. Nucl. Data. Tables }}

\def\PR{{\em Phys. Rev. C }}
\def\PRP{{\em Phys. Rep. }}
\def\PRL{{\em Phys. Rev. Lett. }}

\def\EPJ{{\em Eur. Phys. J. A }}
\def\IJME{{\em Int. J. Mod. Phys. E  }}

\begin{document}
\title{Importance of $\alpha$-induced reactions and the inverse $(\gamma,\alpha)$ on $p$-nuclei}

\author{Chirashree Lahiri\footnote{Email:chirashree.lahiri@gmail.com}\\
Department of Physics, \\Indian Institute of Engineering Science and Technology\\ (Formerly,
Bengal Engineering and Science University),\\ Shibpur, Howrah-711103, India.}

\maketitle

\begin{abstract}
We have calculated astrophysical reaction cross-sections for $(\gamma,\alpha)$ reactions of some nuclei important for the calculation of $p$-process reaction-decay network. Reaction rates for $\alpha$-induced reactions are calculated with the semi-microscopic optical potential constructed using double folding method, where nuclear density distributions for finite nuclei along with the effective nucleon-nucleon interaction are the important components of the folded potential. For this purpose density distributions of target nuclei are obtained   
from relativistic mean field approach. Astrophysical reaction cross section for elastic scattering of $\alpha$-particle from $^{92}$Mo target is compared with the existing experimental results to constrain the newly formed potential. Further, to check the credibility of the present theoretical framework, the astrophysical S-factor for ($\alpha$,$\gamma$) reactions are compared with the experimental observation, wherever available. Finally, an estimate of dominant photodisintegration channels at various astrophysical temperature is discussed for $p$-nuclei  $^{74}$Se and  $^{96}$Ru.

\end{abstract}

\section{Introduction}

With the early development of the theory of nucleosynthesis beyond the iron core, mainly 
two types of nucleosynthesis processes, namely $s$(slow) and $r$(rapid) neutron capture, are identified which can answer the observational foundation of the solar system composition. However,  apart from these two processes, some nuclei are found in nature that can not be produced directly via $s$- or $r$- process ($\sim$ 35 nuclei ranging between $^{74}$Se to $^{196}$Hg found on the proton-rich 
side of the nuclear landscape) are commonly termed as $p$-nuclei\cite{book,arn} . The $p$-process essentially includes 
photodisintegrations of $(\gamma, n)$, $(\gamma, p)$ and/or $(\gamma, \alpha)$ types, along with the captures of neutrons, 
protons and light-particles where centre-of-mass energies typically lie below 1 MeV or the Coulomb barrier in the case of charged particles. 
In order to reproduce the solar system $p$-abundances, the astrophysical environment necessitates a temperature range $\sim$ 2-3 GK, density $\sim$ 10$^6$ gm/cm$^3$, and the time scale  $\sim$ seconds. One of the possible sites of $p$-process that  fulfills these requirements is Type-II supernovae, where this process is expected to develop in the O - Ne layers of the massive stars when the temperature lies between 1.7 and 3.3 GK\cite{arn} . Other anticipated sources involve the pre-Type II supernova, Type -IA supernovae etc. However till date, there has been no definite confirmation regarding the proper astrophysical site for the $p$-process nucleosynthesis.  

A trustworthy modeling of the $p$-process involves the consideration of an extended network of some 20,000 reactions linking about 2000 nuclei in the A$<$210 mass range. One major problem in studying the $p$-process is that many of the nuclei involved in the $p$-network are extremely short lived and are not available in our terrestrial laboratories for experimental analysis. Consequently, it is very difficult to track the entire nucleosynthesis network of the $p$-process experimentally. However, present day experimental arrangements with RIB facility are serving in a promising fashion, we are still very distant from having reaction rates at astrophysical energies for all major reactions involved in
the $p$-process and thus, theoretical study is very important for the network calculation of $p$-process.

In 2006, Rapp \etal \cite{rapp} have identified a number of reactions that are very important in the $p$-process calculation. Further,  abundance ratios of  $(\gamma,n)$ and $(\gamma,\alpha)$ for certain nuclei are of a greater importance as these ratios can sometimes address the abundances
 found in meteoritic inclusions\cite{woo,rau1} . However, previous theoretical calculations of $(\gamma,\alpha)$ reactions rely on the statistical model calculations and they contain a large uncertainty\cite{rau1,rau2,rau3} . Therefore, the introduction of a framework which involves a  microscopic view of a nucleus is expected to provide a model with better accuracy. 

The motivation of the present work is to calculate astrophysical reaction rates of $(\gamma,\alpha)$ channel for some nuclei important for the calculation of the astrophysical $p$-process network. Alongside, astrophysical reaction rates for other photodisintegration processes, viz. $(\gamma,p)$, $(\gamma,n)$ for those nuclei have been calculated. We find it interesting to discuss the dominant decay channels for selected elements in different astrophysical temperatures from the present study.

In the present work, in order to pursue the study of astrophysical reactions  for the $\alpha$-particle plus nucleus system
, semi-microscopic optical model calculations have been performed where nuclear density distributions of the participating nuclei (in this case, the target and the $\alpha$-particle) and the nucleon-nucleon(NN) interactions are the two major components of the calculation. Similar calculations had been implemented successfully to calculate single folded potentials (proton-nucleus potential, neutron-nucleus potential etc.) in our earlier works\cite{clah1,clah2,clah3,clah4} and in Dutta \etal\cite{saumi}.

\section{Methodology}
The semi-microscopic optical potential $V(E,\vec R)$ for an $\alpha$-particle induced reaction 
is calculated by a double folding procedure given by

\begin{equation}
 V(E,\vec R)=\int {d^3r_1}\int {d^3r_2}~\rho(\vec {r_1})\rho(\vec {r_2})v_{eff}(\vec d,E),
\end{equation}

with $\vec d=\vec {r_2}-\vec {r_1}+\vec R$ in fm, where $\vec R$ is the radial separation between the target and the projectile. In the equation, $\rho(\vec {r_1})$ and $\rho(\vec {r_2})$  are the density distributions of the $\alpha$-particle and the target nucleus $X$, respectively for $(X + \alpha)$ reaction. The term $v_{eff}(d,E)$ is the effective NN-interaction obtained either from nuclear matter calculation or from phenomenological models. In the present work, $v_{eff}(d,E)$ have been taken from M3Y interaction defined as

\begin{equation}
v_{eff}(d,E)=7999\frac{e^{-4d}}{4d}-2134\frac{e^{-2.5d}}{2.5d}+J_{00}(E)\delta(d),
\end{equation} 
with the zero range pseudo potential $J_{00}(E)$ given by, 
\begin{equation}
J_{00}(E)=-276\left( 1-0.005\frac{E}{A}\right) {\rm MeV} fm^{3}.
\end{equation}  

Further, theoretical density profile for the nucleus $X$ is extracted from the relativistic mean field (RMF) calculation, which is nothing but the relativistic generalization of the non-relativistic effective theory. This RMF approach has succeeded in explaining different features of stable and exotic nuclei (see P. Ring\cite{ring} and references therein), like ground state binding energy, excited states\cite{shakib,bips} , nuclear deformation etc.   
 and worked in a better way in high density region than the non-relativistic theory. In specific, the radius and the nuclear density are known to be well reproduced which, in turn, lead to its application in the field of nuclear reaction. In RMF,
 there are different variations of the Lagrangian density as well as a number of different parameterizations which are distinct from
each other in various ways (like inclusion of new interaction or different value of
masses and coupling constants of the meson, etc). In the present work we have employed the FSU Gold Lagrangian density\cite{fsu} to calculate the density distribution $\rho(\vec {r_2})$ . This set of parameters have been successfully used in earlier works where semi-microscopic optical potential for proton-nucleus\cite{clah1,clah2,clah3,clah4} and neutron-nucleus\cite{saumi} was calculated using single folding method. 

Astrophysical reaction calculations have been performed with the computer code TALYS 1.8\cite{tal} assuming the target nucleus spherically symmetric. The M3Y interaction is not the standard input option of TALYS and hence we have modified the code by incorporating the interaction. Further, in order to obtain the nucleus-nucleus potential of spherical nuclei by using the double folding model, the code DFPOT\cite{cpc} , modified to accommodate density distribution from RMF calculation, has been used. The inclusion of spin-orbit term in the folded potential has been adopted from Lahiri and Gangopadhyay\cite{clah1, clah2, clah3,clah4}.  

In the field of nuclear astrophysics, while dealing with low energy astrophysical reactions, experimental observation of reaction cross-section in the low energy region is extremely challenging as the cross-section shows a sharp drop with decreasing energy. At the same time, for experimental data evaluated at a relatively high energy, it is extremely difficult to extrapolate them towards the low energy domain. Such extrapolation invites large amount of errors as the reaction cross-section varies very rapidly at low energy.
 In order to avoid this difficulty, usual practice is to calculate another key observable, known as astrophysical S-factor. It can be expressed as\cite{book}

\begin{equation}
S(E)=E\sigma(E)e^{2\pi\eta},
\end{equation}

where $E$ is the energy in center of mass frame ($E_{cm}$) in keV. 
 The Sommerfeld parameter is expressed as,  
\begin{equation}
 2\pi\eta=31.29 Z_{P}Z_{X}\sqrt{\frac{\mu}{E}}.
\end{equation}

Here $\sigma(E)$ is in  barn, $Z_{P}$ and $Z_{X}$ are the charge numbers 
of the projectile and the target, respectively and $\mu$ is the reduced 
mass (in amu) of the composite system. This S-factor varies much slowly 
than reaction cross-sections and therefore, we calculate this quantity
and compare it with experimentally obtained values.   

Further, in equilibrium, the stellar decay constant ($\lambda$) for photodisintegration, i.e. $(\gamma,\alpha)$, $(\gamma,p)$ and $(\gamma,n)$ reactions, are
related to the $\alpha$, proton and neutron-capture reactions, respectively by the reciprocity theorem\cite{book} . As an example, for a reaction(forward) $X+ P \rightarrow Y+\gamma$, where $P$ is the projectile, the expression has a form

\begin{eqnarray}
\lambda=
9.86851\times10^9 T^{\frac{3}{2}}\left(\frac{M_PM_{X}}{M_{Y}}\right)^{\frac{3}{2}}
\frac{\left( 2J_{P}+1 \right)\left( 2J_{X}+1 \right)}
{\left( 2J_{Y}+1 \right)}\frac{G_P G_X}{G_Y}
\nonumber\\
{N\langle\sigma v\rangle_{P X\rightarrow Y\ga}} 
\exp\left(\frac{-11.605Q}{T}\right),
\end{eqnarray}

in the unit of $sec^{-1}$, where forward reaction rate $(P,\gamma)$ is expressed in 
$cm^3mol^{-1}sec^{-1}$. Here, in Eq. (6), temperature $T$ is in GK(10$^9$ K), Q(MeV) is the reaction Q value of the forward reaction adopted from NNDC\cite{nndc} , $G_X$ and $G_Y$ are the normalized partition functions for the target $X$ and residual $Y$, respectively from Rauscher and  Thielemann\cite{rau3}. For the projectile P, we choose $G_P$=1.

\section{Results and Discussions}

In order to check the credibility of the present semi-microscopic potential, we have calculated elastic $\alpha$-scattering cross section
where experimental data are available. As the elastic scattering process involves
the same incoming and outgoing channel for the optical model, it is expected to
provide the simplest way to constrain various parameters involved in the calculation.

In a $p$-process calculation with Z $\sim$ 30 - 80 and A $\sim$ 70 - 200, while considering $\alpha$- particle induced reactions, astrophysically important Gamow peak energies lie roughly in the ballpark of $\sim$ 5 - 12 MeV when temperatures of the environment is $\sim$ 2 - 3 GK. For example, the Gamow peak energy for a $^{92}$Mo+$\alpha$ reaction lies in between 5.8 - 7.6 MeV, whereas for a reaction $^{151}$Eu+$\alpha$, the peak energy lies within 7.7 - 10.0 MeV range in the  temperature window mentioned above.    

 However, scattering experiments are very difficult at such low energies
because of extremely small reaction cross-sections, and hence no experimental data are available in that region. Therefore, cross-sections from our calculations have been compared with the lowest energy experimental data available in the literature.

\begin{figure}[ht]
\center
\includegraphics[scale=0.6]{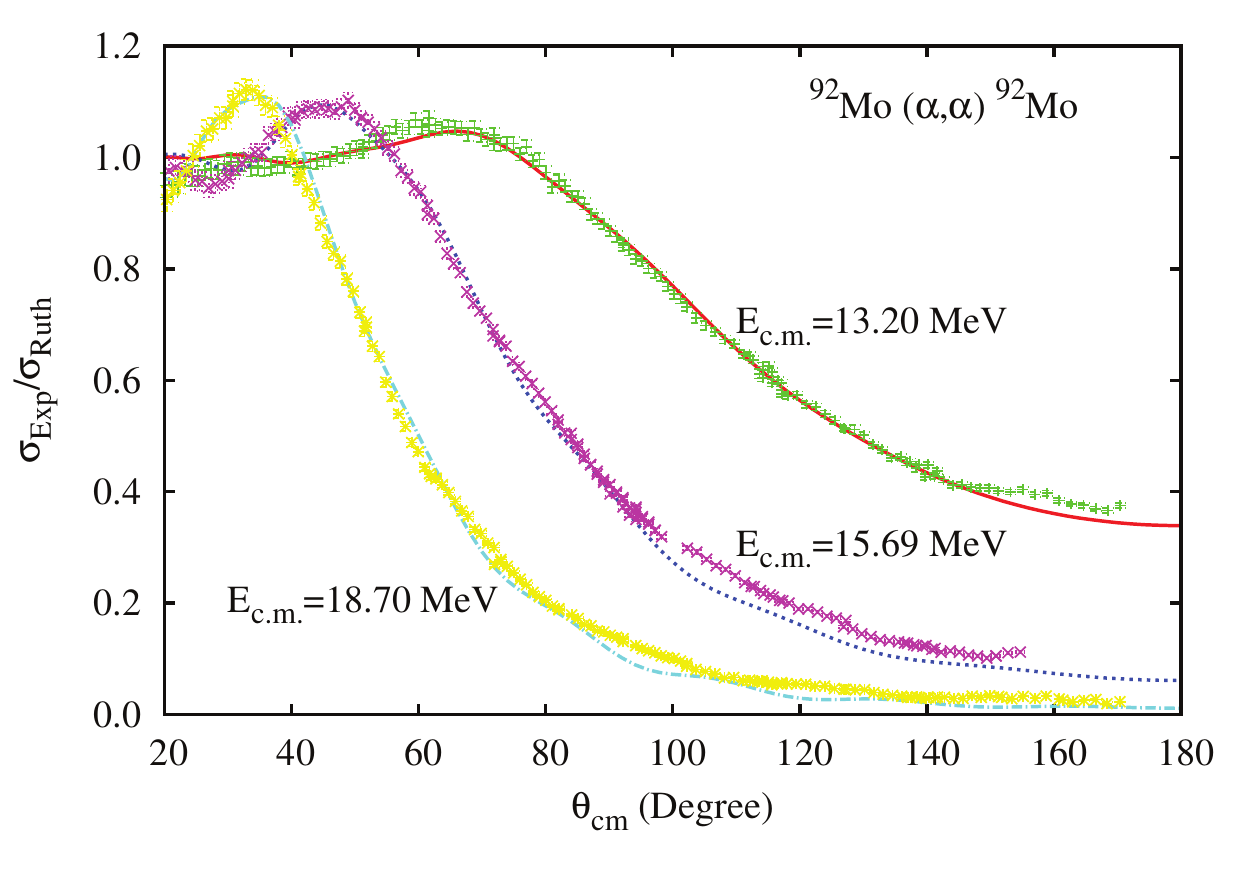}
\caption{Elastic scattering cross sections from our calculation normalized to the Ruther-ford cross section
is compared with experimental data\cite{mo92}\label{elastic}.}
\end{figure}

In Fig. \ref{elastic}, the results for the elastic scattering of $^{92}$Mo$(\alpha,\alpha)^{92}$Mo is presented with
the corresponding experimental results\cite{mo92} . In order to fit the experimental data, the double folded 
potential is multiplied by factors 0.7 and 0.3 to obtain the
real and the imaginary parts of the optical potential, respectively. It should be noted that better fits for individual
reactions can be achieved by varying different parameters, but this approach is not felicitous if
the present calculation has to be extended to explore unknown nuclei/mass region.
Therefore we have used a single parameterization using these two factors, throughout the rest of this work.

\begin{figure}[h]
\center
\includegraphics[scale=0.6]{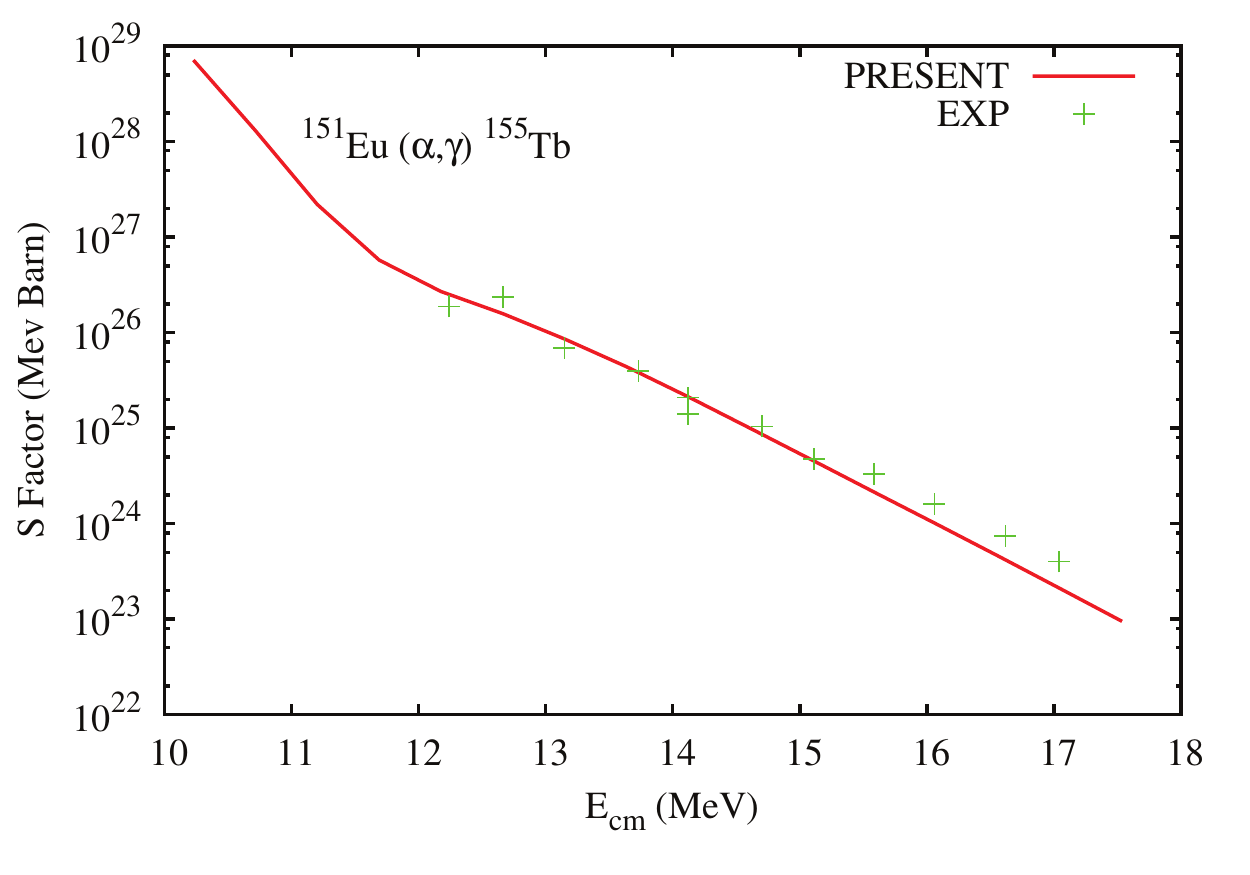}
\caption{Astrophysical S-factor from our calculation for $^{151}$Eu$(\alpha,\gamma)^{155}$Tb compared with the experimental data from Gy\"urky \etal \cite{eu151}\label{sf}.}
\end{figure}

In case of $p$-process nucleosynthesis, $^{151}$Eu$(\alpha,\gamma)^{155}$Tb is one of the important reactions\cite{eu151} in the $p$-network around the Eu-Gd-Tb region.  
In Fig. \ref{sf}, calculation of the astrophysical S-factor from our calculation for the above mentioned reaction is compared to the available experimental data\cite{eu151} . It is clearly visible from Fig. \ref{sf} that the current calculation matches with the experimental data in a good manner. The agreement ensures the credibility of our present theoretical model and thus enables us in employing this model in the region where experimental values are unavailable. 

In Rapp \etal\cite{rapp} , the authors have identified a few reactions that, together with their respective
inverse reactions, have been found to exhibit the strongest influence on
the final $p$-abundances. Particularly, the $(\gamma,\alpha)$ reaction for $^{74}$Se and $^{96}$Ru nuclei are identified as important landmarks in the $p$-process scenario. 

\begin{figure}[h]
\center
\includegraphics[scale=0.6]{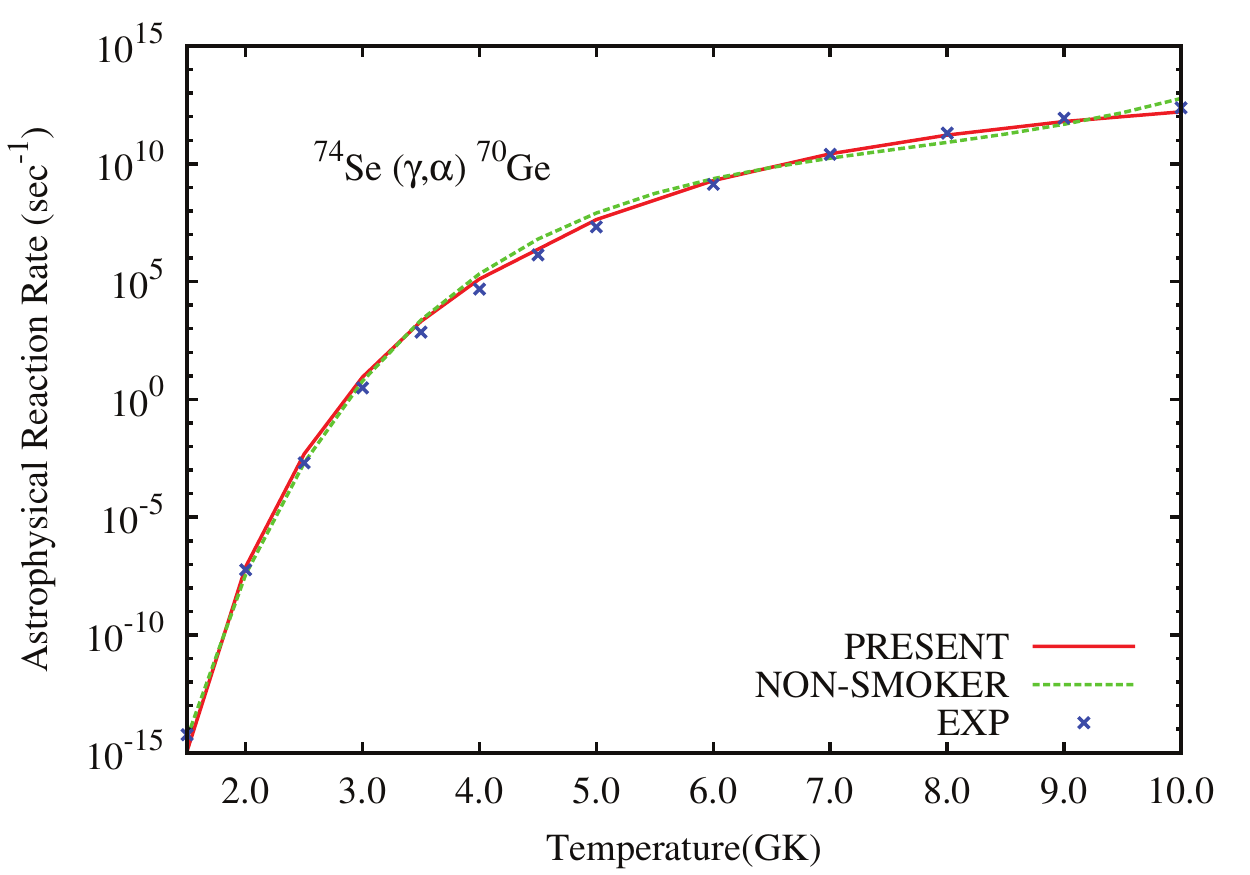}
\caption{Astrophysical reaction rate from our calculation for $^{74}$Se$(\gamma,\alpha)^{70}$Ge compared with the experimental data from F\"ul\"op \etal \cite{se74} and earlier theoretical study\cite{rau3} using NON-SMOKER code\cite{nsmo}\label{se}.}
\end{figure}

In Fig. \ref{se}, astrophysical reaction rates for $^{74}$Se$(\gamma,\alpha)^{70}$Ge reaction from our present calculation are compared with the previously available experimental, as well as theoretical predictions. In 1996, F\"ul\"op \etal\cite{se74} have studied the $^{70}$Ge$(\alpha,\gamma)^{74}$Se reaction using single and coincidence gamma spectroscopy techniques. The authors have provided the inverse rate, denoted as `EXP' in Fig. \ref{se}. The dataset `NON-SMOKER' in Fig. \ref{se} is obtained by using Eq. (6), which involves
the statistical model calculation from NON-SMOKER code\cite{rau1,rau2,rau3,nsmo} for $\alpha + ^{70}$Ge as the forward reaction. It is visible from the figure that there is a nice agreement of our present calculation with the previous NON-SMOKER calculation. Further, one can see that the present calculation matches the experimental data\cite{se74} in a slightly better fashion than the NON-SMOKER calculation. 

\begin{figure}[h]
\center
\includegraphics[scale=0.6]{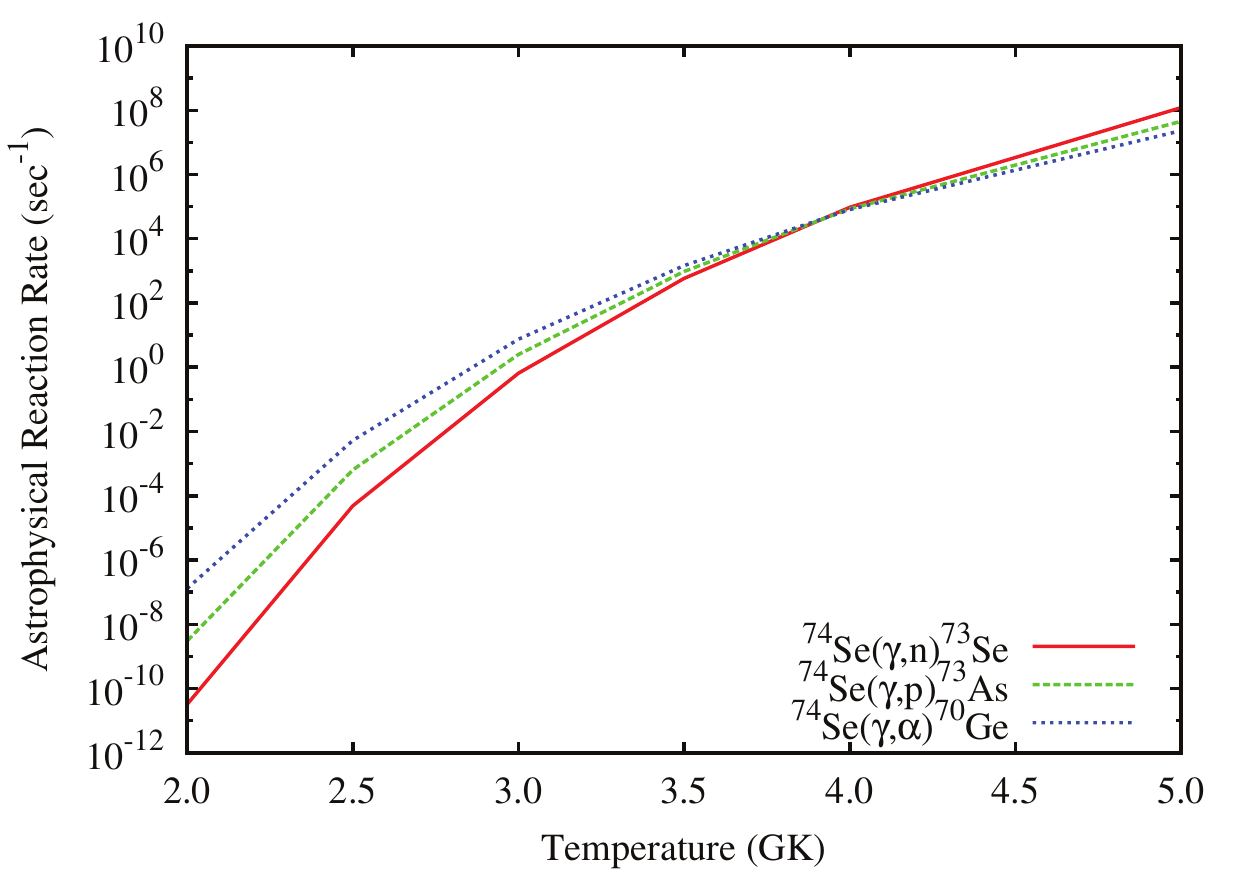}
\includegraphics[scale=0.6]{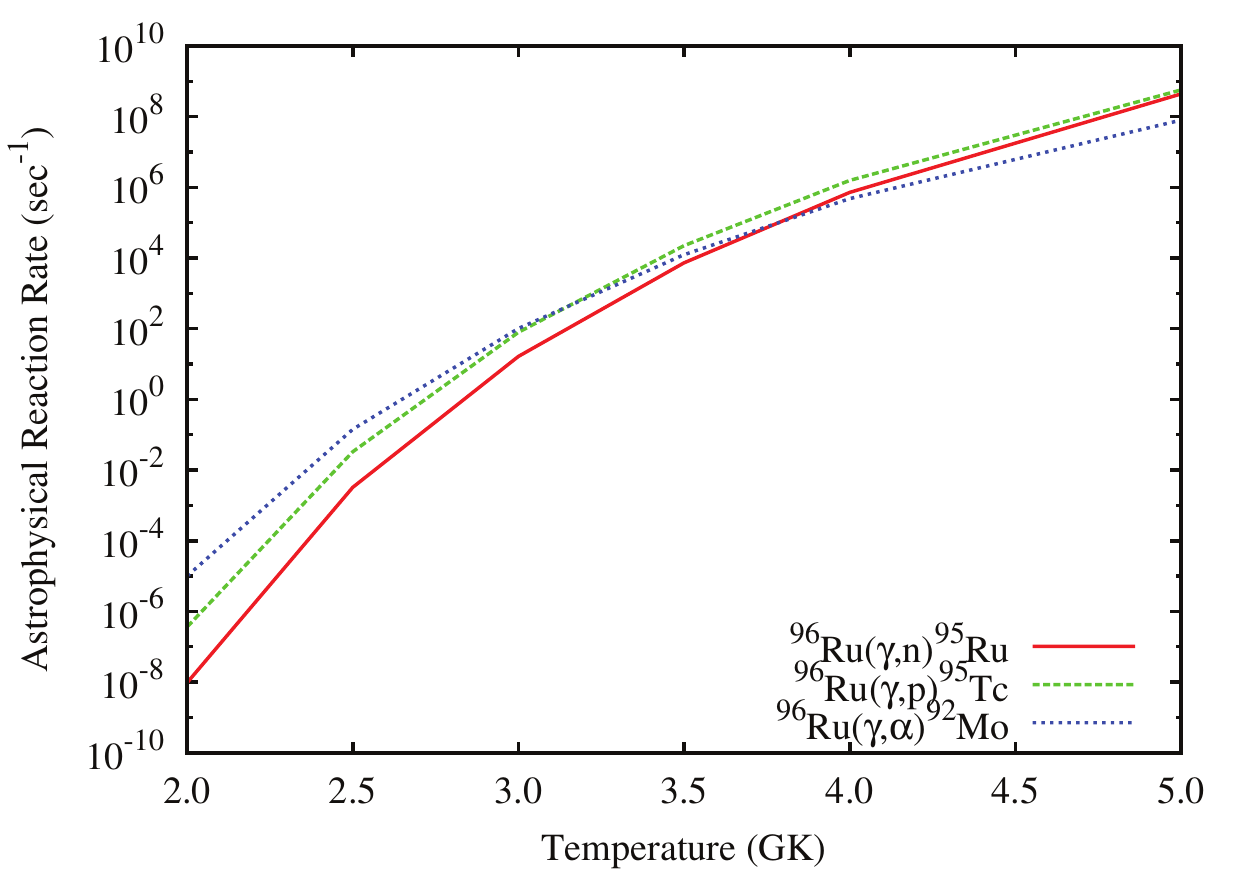}
\caption{Variation of astrophysical reaction rates for $(\gamma,n)$, $(\gamma,p)$ and $(\gamma,\alpha)$ reactions with temperature for $^{74}$Se (above) and $^{96}$Ru(below) nuclei\label{comp}.}
\end{figure}

In Fig. \ref{comp}, variation of astrophysical reaction rates for $(\gamma,n)$, $(\gamma,p)$ and $(\gamma,\alpha)$ reactions with the temperature are shown for $^{74}$Se and $^{96}$Ru nuclei. In the figure, the dashed line (green) denotes the rate for  $(\gamma,p)$ reaction taken from our previous work\cite{clah1}. The continuous (red) line is for $(\gamma,n)$ reaction, calculated with Eq. (6)  from $(n,\gamma)$ reaction rate which is, in turn, obtained by using single-folded semi-microscopic optical potential\cite{clah1,clah2,clah3} . One can see that $(\gamma,\alpha)$ reaction, denoted by the dotted (blue) line in Fig. \ref{comp}, dominates over other photodisintegration channels in the low temperature. For example, in case of $^{74}$Se nucleus,  $(\gamma,\alpha)$ reaction rate is greater by $\sim$ 10$^2$ times  from $(\gamma,p)$ rate and $\sim$ 10$^4$ times from $(\gamma,n)$ rate around 2 GK temperature. The difference between these three rate values decrease with increasing temperature and the situation changes around 4 GK temperature, where $(\gamma,n)$ and $(\gamma,p)$ reactions start dominating over the $(\gamma,\alpha)$ channel.
Therefore, from Fig. \ref{comp} (above panel), it can be inferred that the decay of $^{74}$Se mainly leads to the production of $^{70}$Ge in the temperature range 2-4 GK, which is the requisite temperature window for the $p$-process nucleosynthesis as per our present day knowledge. Clearly, in the decay of $^{74}$Se, the elements  $^{73}$As and $^{73}$Se remain under-produced compared to the production of $^{70}$Ge at a temperature $<$ 4 GK. However, the scenario changes around the temperature 4 GK as the probability of decay to all three competing channels are of the comparable order at the mentioned temperature. As a result, rather than choosing the path only via $\alpha$- decay, the $p$-process path from $^{74}$Se trifurcates around 4 GK as the mass accumulation of $^{73}$Se and $^{73}$As nuclei become higher at this temperature  due to the enhancement in $(\gamma,n)$ and $(\gamma,p)$ reaction rates, respectively.

 Similar scenario takes place for photodisintegration processes from $^{96}$Ru. In this case, it is visible from Fig. \ref{comp} (lower panel) that  $(\gamma,\alpha)$ rate dominates at low temperature similar as that of the  $^{74}$Se case. Around 3 GK, $(\gamma,p)$ leads the competition, whereas $(\gamma,n)$ lags behind. Finally, $(\gamma,n)$ starts to dominate above the 4 GK temperature for the decay of $^{96}$Ru.

It is clearly evident from Fig. \ref{comp} that the dependence of photodisintegration rates to various decay channels is highly sensitive to the temperature. Therefore, it can be stated that the final fate of a nucleus in the $p$-process network depends mainly on the temperature of the astrophysical environment. However, it is important to mention that the temperature where two or more photodisintegration rates become almost comparable, i.e. where the rate curves intersect, (for example, for $^{74}$Se around 4 GK ) is not unique for all nuclei. In fact, there are several numbers of $p$-nuclei, where only one of the photodisintegration channels [example: $(\gamma,\alpha)$ channel for $^{196}$Pb; $(\gamma,p)$ channel for $^{100}$Pd] pay the accountable contribution and as a result no cross-over temperature exists in the concerned temperature window for those nuclei. 

For the sake of completeness, we have re-visited the results from Fig. \ref{comp} using a different set of RMF parameters, namely NL3\cite{clah4,nl3} . It is observed that the results for reaction rates with NL3 parameters mostly differ in the second decimal place and effectively has no notable deviation from the results of Fig. \ref{comp}. The observation suggests that the findings of the Fig. \ref{comp} are independent of the choice of RMF parameterizations.      

\section{Summary}
To summarize, the present work mainly deals with the impact of $(\alpha,\gamma)$ and its reverse process on some astrophysically important $p$-nuclei. The theoretical calculation of astrophysical reaction for $\alpha$-capture is performed using semi-microscopic optical model, where phenomenological M3Y interaction is folded with density distributions of the projectile and the target nuclei by the method of double folding. In order to calculate the density distribution of the target nucleus, relativistic mean field Lagrangian density FSU-Gold has been used. The model parameters for this potential have been regulated by comparing our results to the available experimental data for elastic scattering cross-section. It is found that the present theoretical modelling is successful in reproducing the experimental S-factor for the $\alpha+^{151}$Eu reaction, an important reaction in the $p$-network. In the next step, astrophysical reaction rates are calculated for the inverse process, i.e. $(\gamma,\alpha)$. Finally, a comparative study of all photodisintegration processes, involved in the $p$-process nucleosynthesis have been performed for $p$-nuclei $^{74}$Se and $^{96}$Ru. The study give us an idea about the temperature dependence of the competing reaction channels and we find that the $(\gamma,\alpha)$ channel dominates mainly in the low temperature region. In addition, we conclude that our present observation is independent for different choices of RMF parameterizations.

\section*{Acknowledgment}

The author acknowledges the grant from DST-NPDF (No.PDF/2016/001348) Fellowship.

\end{document}